\documentclass[prd,twocolumn]{revtex4}
\usepackage{graphicx}
\usepackage{amssymb}

\begin{document}

\title{{\bf Gauge Coupling Variation in Brane Models}}

\author{G. A. Palma$^{\, 1}$, Ph. Brax$^{\, 2}$, A. C. Davis$^{\, 1}$ and C. van de Bruck$^{\, 3}$}

\affiliation{$^{1}$Department of Applied Mathematics and
Theoretical Physics,
Center for Mathematical Sciences, University of Cambridge, Wilberforce Road, Cambridge CB3 0WA, U.K. \\
$^{2}$Service de Physique Theorique, CEA-Saclay F-91191 Gif/Yvette Cedex, France. \\
$^{3}$Astrophysics Group, University of Oxford, Denys Wilkinson
Building, Keble Road, Oxford OX1 3RG U.K.}


\begin{abstract}
We consider the space-time variation of gauge couplings in
brane-world models induced by the coupling to a bulk scalar field.
A variation is  generated by the running of the gauge couplings
with energy and a conformal anomaly while going from the Jordan to
the Einstein frame. We indicate that the one-loop corrections
cancel implying that one obtains a variation of the fine structure
constant by either directly coupling the gauge fields to the bulk
scalar field or having bulk scalar field dependent Yukawa
couplings. Taking into account the cosmological dynamics of the
bulk scalar field, we constrain the strength of the gauge coupling
dependence on the bulk scalar field and relate it to modifications
of gravity at low energy.
\end{abstract}


\maketitle

Physics beyond the standard model predicts the existence of scalar
degrees of freedom, moduli fields, whose presence modify general
relativity even at low energy. Within the framework of brane-world
models these moduli fields are directly associated with the
possible deformations of the bulk-brane configurations, i.e. the
free motion of branes relative to each other (see e.g.
\cite{braxvdbruckreview} for recent reviews on brane worlds). The
presence of such massless fields is generally thought to lead to
large deviations with respect to general relativity, prompting the
need for a stabilization of the moduli. Within string theory this
is the case for the dilaton field whose expectation value is
related to the gauge coupling constant at the unification scale.
Such a running dilaton is ruled out experimentally. Attempts to
stabilize the dilaton have been notoriously difficult to justify
and require ingredients such as non-perturbative effects and
supersymmetry breaking.

The time evolution of the moduli fields does not only imply
modifications to general relativity but may also provide a hint
towards extensions to the standard model sector. Phenomena such as
the variation of coupling constants may spring from the fact that
these couplings are effective parameters  depending  on the
moduli. In this paper we will discuss the time-variation of
coupling parameters in brane models. (See e.g. \cite{Calmet &
Fritzsch}--\cite{Kostelecky} for recent discussions on varying
constants. Reference \cite{uzan} is a recent discussion on
experimental and theoretical aspects of varying constants and has
an extensive list of literature.)

We restrict the present analysis to the case in which the gauge
couplings depend on the evolution of a single scalar field $\psi$.
Let us consider the gauge couplings $\alpha_{i}=g_i^2/4\pi$ to one
loop order
\begin{eqnarray} \label{eq: running 1}
\alpha_{i}^{-1}(E,\psi ) = \alpha_{i}^{-1}(\tilde E,\psi ) + \frac{b_{i}}{2
\pi} \ln (\tilde E / E),
\end{eqnarray}
where $E$ and $\tilde E$ are two arbitrary energy scales, and
$b_{i}$ are the renormalization group equation coefficients.
Notice that the renormalization group coefficients are pure
numbers determined solely by the matter content of the model.
These coefficients may be taken as the standard model ones,
$b^{SM}_{i} = (41/10, -19/6, -7)$. We have emphasized the explicit
dependence on the scalar field $\psi$. In this way, a variation of
the couplings $\alpha_{i}$ over cosmological meaningful periods
would be a consequence of the evolution of $\psi$.

In the Einstein frame where the Planck mass is time-independent,
the physical masses $M$ may depend on the scalar field $\psi$. If
this is the case, the total variation of the gauge coupling
involves the  variation of $M$. From relation (\ref{eq: running
1}), this total variation reads
\begin{eqnarray} \label{eq: var M}
\delta \alpha_{i }^{-1}(M,\psi) = \delta_\psi
\alpha_{i}^{-1}(m_{\rm Pl}, \psi) - \frac{b_{i}}{2 \pi}
\frac{\delta_\psi M}{M},
\end{eqnarray}
where $M$ is the energy scale at which one measures the gauge
coupling. One can be more specific in a large class of theories
where the effective four dimensional action has the generic form
\begin{eqnarray} \label{eq: Jordan}
S_{\rm J} \supset \int d^4x \sqrt{-g}\left[
\frac{A^{-2}{\cal R}}{{2\kappa^2_4}} - \frac{1}{4g_{J}^2} F^2 \right],
\end{eqnarray}
in the Jordan frame where matter couples to $g_{\mu\nu}$, as
opposed to the Einstein frame, where  Newton's constant is moduli
field-independent. In the above expression $A$ depends on the
scalar field $\psi$, $F$ is the antisymmetric gauge tensor field,
and $g_{J}$ is the gauge coupling in the Jordan frame (which could
have a dependence on the moduli field). To go to the Einstein
frame formulation we must perform the conformal transformation
\begin{eqnarray} \label{eq: g - Ag}
g_{\mu \nu} \rightarrow  A^{-2} g_{\mu \nu}.
\end{eqnarray}
The masses of particles are affected according to \cite{US}
\begin{equation} M_{E} = A M_{J},
\end{equation}
while the bare dimensionless coupling constants remain
unaffected, $g_{J} \rightarrow g_{E} = g_{J}$. This picture is
modified quantum mechanically \cite{Fujii Maeda}. In the Einstein
frame, the relation between the bare and the renormalized
couplings, $g_B$ and $g$, using dimensional regularization is
\begin{equation}
g_B = \mu^{2 - d/2} g \left[ 1 + g^2 \frac{b}{32 \pi^2}
\Gamma(2 - d/2)  \right],
\end{equation}
where $\mu$ is the renormalization group energy scale.
The dependence between the bare coupling $g_{B}$ in the Einstein
frame and the bare coupling $g_{J}$ in the Jordan frame is $g_{B}
= A^{2-d/2} g_{J}$. Then, it follows
\begin{equation}
g_J = (\mu/A)^{2 - d/2} g \left[ 1 + g^2  \frac{b}{32 \pi^2}
\Gamma(2 - d/2)  \right].
\end{equation}
We consider its direct dependence on the scalar field $\psi$ by
introducing a function $D(\psi)$ in such a way that
\begin{equation}
4 \pi \, \delta_\psi (g_{J}^{-2}) = \delta_\psi D(\psi).
\end{equation}
Now, taking into account the dependence of both the scale $\mu=M$ and $A$ on
$\psi$ we find that in the limit $d=4$
\begin{equation}
\delta g = \beta (g) \left[ \frac{\delta_\psi M}{M} -
\frac{\delta_\psi A}{A} \right] - \frac{g^3}{8 \pi} \delta_\psi D(\psi),
\end{equation}
where $\beta (g) = b \, g^{3}/16 \pi^2$ is the renormalization
beta function. In terms of $\alpha$, the above variation can be
written as
\begin{equation}
\delta \alpha^{-1} = \frac{b}{2 \pi} \left[ \frac{\delta_\psi A}{A} -
\frac{\delta_\psi M}{M} \right] + \delta_\psi D(\psi).
\end{equation}
Thus, if $M \sim A$, the variation of the gauge coupling is purely
due to its direct dependence on $\psi$. In general, one expects
that masses of particles appear due to the Higgs mechanism where
$M_J=\lambda (\psi)v$. Here $\lambda (\psi)$ is a Yukawa coupling
which may depend on $\psi$ and $v$ the Higgs expectation value. In
the Einstein frame this leads to
\begin{equation} \label{final}
\delta \alpha^{-1} = - \frac{b}{2 \pi}  \frac{\delta_\psi
\lambda}{\lambda}+ \delta_\psi D(\psi), \label{frame}
\end{equation}
showing that the variation of gauge coupling constants picks up a
one loop contribution from the scalar field dependence of the
Yukawa couplings. Let us emphasize that this result is the same
one in the Jordan frame, where the conformal factor $A$ is absent
in the matter sector. In fact the result (\ref{frame}) is
frame-independent. In the following we shall assume that Yukawa
couplings do not depend on $\psi$. Additionally, if we assume
grand unification, then we are forced to take $D(\psi)$ as
independent of the gauge sector $i$. From now on we consider that
this is the case.

We now focus on the confinement mass scale $\Lambda_{QCD}$
for the strong sector of the theory. This mass scale is defined by
$\alpha_{3}^{-1}(E) = -
(b_{3} / 2 \pi) \ln ( E / \Lambda_{\mathrm{QCD}})$
implying that it depends on $\psi$
\begin{eqnarray}\label{QCD}
\Lambda_{QCD} \sim A(\psi) \exp \left[ \frac{2 \pi}{b^{SM}_{3}}
D(\psi) \right],
\end{eqnarray}
in the Einstein frame.
This relation turns out to be highly relevant  since, in the chiral
limit, the masses of baryons are proportional to the QCD
confinement mass scale.

Let us now turn to the variation of the fine structure constant.
We can use its dependence on the electroweak couplings
$\alpha_{1}$ and $\alpha_{2}$: $\alpha_{EM} = \frac{3}{5} \cos^{2}
(\theta_{W}) \alpha_{1}$, and $\alpha_{EM} = \sin^{2} (\theta_{W})
\alpha_{2}$, where $\theta_{W}$ is the weak mixing angle.
In general,  $\theta_{W}$ is also dependent on $\psi$.
The  relation between $\alpha_{1}$, $\alpha_{2}$
 and the fine structure $\alpha_{EM}$ at a fixed energy
scale $E$ is  $\delta \alpha_{EM}^{-1} (E)  = (5/3)
\delta \alpha_{1}^{-1} (E) + \delta \alpha_{2}^{-1} (E)$. In this
way, the total variation of the fine structure constant is
\begin{eqnarray} \label{eq: fine var}
\frac{\delta \alpha_{EM}}{\alpha_{EM}} = -  \, \alpha_{EM}
\frac{8}{3} \, \frac{\partial D}{\partial \psi} \, \delta \psi .
\end{eqnarray}

Another measurable quantity intimately related to the former
variation is the ratio of the proton mass $m_{p}$ to the electron
mass $m_{e}$, $\mu = m_{p} / m_{e}$. Since in our model the
electron mass has the behaviour $m_{e} \sim A$, we obtain $\mu =
\exp \left[ (2 \pi / b^{SM}_{3}) D(\psi) \right]$, where we have
used (\ref{QCD}). Therefore, in the class of theories which we are
considering, a variation of $\mu$ can be cast into $\delta \mu /
\mu = (2 \pi / b^{SM}_{3}) \, \delta D(\psi)$, or by directly
relating it to the variation of $\alpha_{EM}$, as
\begin{eqnarray} \label{eq: mu var}
\frac{\delta \mu}{\mu} = - \frac{3}{8 \, \alpha_{EM}} \, \frac{2
\pi}{b^{SM}_{3}} \, \frac{\delta \alpha_{EM}}{\alpha_{EM}} .
\end{eqnarray}
Since $b^{SM}_{3}$ is negative, the variation of $\mu$ has the
same sign as  the variation of $\alpha_{EM}$. In the above
expressions, we will use $b^{SM}_{3} = -9$, which is the value
corresponding to three light quark flavours.


We now apply the above results to brane-world models with a bulk
scalar field and two boundary branes. The bulk scalar field will
take the role of $\psi$ in the discussion above. More particularly
we focus on BPS configurations where the two boundary branes are
free to move without hindrance \cite{Davis & Brax}. This results
in the presence of two massless moduli fields at low energy whose
dynamics have been described in a low energy effective action
obtained after integrating over the fifth dimension \cite{US}. The
effective theory is of the tensor-scalar form leading to
corrections to general relativity. Cosmologically the moduli
fields are coupled to pressureless matter leading to a time
dependence in the matter dominated era.  In the Einstein frame,
the action reads
\begin{eqnarray}
S_{\rm EF} = \frac{1}{2\kappa_4^2} \! \int \!\! d^4x \sqrt{-g} \!
\left[ {\cal R} -  3 \lambda \bigg( \alpha^2 \, \partial \phi^2 +
\frac{1}{4}
 \, \partial R^2 \bigg) \right] ,
\end{eqnarray}
where $\phi$ and $R$ are the two (unnormalized) moduli fields. The
constant $\alpha$ springs from the coupling of the bulk scalar
field $\psi$ to the branes. More precisely, the coupling of the
scalar field to the brane reads $\int d^4x \sqrt{-g_4} \, T(\psi)$
where $T(\psi)$ is a bulk scalar dependent function, $g_4$ is the
induced metric on the brane and we identify $\delta T/T = \alpha
\, \delta \psi$ evaluated at the present time. When $\alpha=0$ we
retrieve the Randall-Sundrum model with no bulk scalar field. For
larger values of $\alpha=O(1)$ one obtains the low-energy
effective action of heterotic $M$ theory, taking only into account
the volume of the Calabi-Yau manifold \cite{Lukas}. The low energy
effective action comprises the Einstein-Hilbert term and the
kinetic terms of two massless fields $\phi$ and $R$. In addition
we will consider a gauge kinetic term. The coupling to ordinary
matter is field dependent $S_m^{(1)} =
S_m^{(1)}[\Psi_1,A^2(\phi,R)g_{\mu\nu}]$ and $S_m^{(2)} =
S_m^{(2)}[\Psi_2,B^2(\phi,R)g_{\mu\nu}]$, where we have
distinguished matter living on the positive tension brane labelled
(1) and the negative tension brane labelled (2). The coupling
constants $A$ and $ B$ are given by \cite{US} $A=e^{-
\alpha^2\lambda/2 \, \phi}(\cosh R)^{\lambda/4}$ and
$B=e^{-\alpha^2\lambda/2 \, \phi}(\sinh R)^{\lambda/4}$, where
$\lambda=4/(1+2\alpha^2)$. One  can write the action in the Jordan
frame of the positive tension brane, i.e. in terms of the induced
metric $g^B_{\mu\nu}=A^2 g_{\mu\nu}$. The new action comprises the
Einstein-Hilbert term and the gauge boson kinetic terms, having
the same form as (\ref{eq: Jordan}), with the same dimensionless
factor $A$.

We have already seen  that when the moduli fields manifest
themselves only through the dimensionless  factor $A$ of (\ref{eq:
Jordan}) the variation of the gauge coupling constants is due to
\begin{equation}\label{ansatz}
\delta_\psi D(\psi)=\beta \delta\psi,
\end{equation}
where $\beta=\partial D/\partial \psi \vert_{\psi_0}$ and $\psi_0$
is the present day value of $\psi$. In this expression we have
neglected second-order terms in $\delta \psi$. The dependence of
the scalar field $\psi$ in terms of the moduli fields is given by
\cite{US} $\psi = - \alpha \lambda (\phi + \ln [\cosh R])$.

Brane-world models lead to  modifications to general relativity.
Recalling that baryons have the dependence $\tilde m \sim
\Lambda_{QCD}$, and defining $\alpha_{\phi} = \partial_{\phi} \ln
\tilde m$ and $\alpha_{R} = \partial_{R} \ln \tilde m$, we obtain:
\begin{eqnarray}
\alpha_{\phi} &=& - \alpha \lambda \left[ \frac{2 \pi}{b^{SM}_{3}}
\beta +
\frac{\alpha}{2} \right], \label{eq: alpha phi} \\
\alpha_{R} &=&  - \lambda \left[ \alpha  \frac{2 \pi}{b^{SM}_{3}}
\beta - \frac{1}{4} \right] \tanh R.
\end{eqnarray}
As already said, the variation of gauge coupling constants is
related to the corrections to general relativity as measured in
the solar system. In particular the post-Newtonian Eddington
coefficient $\gamma$ is constrained by the very long baseline
interferometry measurements of the deflection of radio waves by
the sum to be $ -3.8 \times 10^{-4}<\gamma  -1 <2.6 \times
10^{-4}$ at $68 \%$ confidence level \cite{eubanks}. This is
related to the parameter $\theta$ by $\gamma -1\approx -2\theta$,
where
\begin{equation}
\theta= \frac{1+2\alpha^2}{12\alpha^2}\alpha_\phi^2 +
\frac{2\alpha^2+1}{6}\alpha_R^2 .
\end{equation}
The parameter $\theta$ is a measure of the coupling of the moduli
to  matter which will be taken to be $\theta \lesssim 10^{-4}$
\cite{Will}.

\begin{figure}[ht]
\begin{center}
\includegraphics[width=0.47\textwidth]{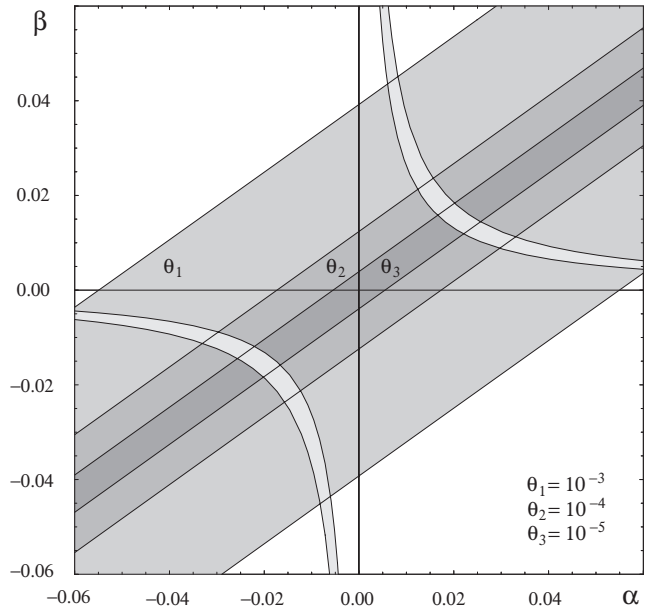}
\caption[Constraints on $\alpha$ and $\beta$]{Constraints on the
parameters $\alpha$ and $\beta$ for three different values of
$\theta$. These values are $\theta_{1}= 10^{-3}$, $\theta_{2}=
10^{-4}$ and $\theta_{3}= 10^{-5}$. The constrained region for
$\alpha$ and $\beta$ is restricted by the contour curves
parameterized by $\theta$ and the hyperbolic curves denoting the
favoured region for a variation of the fine structure according to
Webb {\it et al.}.} \label{F1}
\end{center}
\end{figure}

We now turn our attention to the variation of the fine structure
constant  $\alpha_{EM}$. Using the known cosmological evolution of
the moduli fields in the matter dominated era, we can infer the
variation of the fine structure constant in the recent past. It
was found in \cite{US}, that the field $R$ decays quickly during
the matter dominated era and is small in the redshift range
between $z=0$ and $z=5$. Therefore, we can neglect the influence
of $R$. Taking the cosmological variation of $\psi$, in terms of
the moduli $\phi$, as $\delta \psi = - \, \alpha \lambda \, \delta
\phi$  and using the solution for $\psi$ given by $\delta \phi =
\phi (z) - \phi_{0} = -(1/6)\ln (1+z)$ as a function of the
redshift $z$ (where $\phi_{0}$ refers to the present value of
$\phi$) in equation (\ref{eq: fine var}), the resulting expression
for the variation of the fine structure is
\begin{eqnarray}
\frac{\delta \alpha_{EM}}{\alpha_{EM}} =    - \, \alpha_{EM}
\frac{4}{9} \frac{ 4 \alpha \beta}{1 + 2 \alpha^{2}}  \ln (1 + z).
\end{eqnarray}
(Note that we have neglected the era of vacuum domination between
$z=1$ and $z=0$.) Taking the value $\delta \alpha_{EM} /
\alpha_{EM} \approx - 0.6 \times 10^{-5}$ at $z = 3$ from Webb
{\it et al.} \cite{Webb 1}--\cite{Webb 3}, we must then have
$|\alpha \beta| \approx 3 \times 10^{-4}$. Assuming that $\alpha$
and $\beta$ are of the same order, then $\alpha \approx 10^{-2}$,
which is still in the acceptable region given by the $\theta$
constraint shown above. Figure \ref{F1} shows the favoured region
for the parameters $\alpha$ and $\beta$ subject to the $\theta$
constraint and the measured fine structure variation.

According to recent observations, at a redshift of about one, the
expansion of the universe starts to accelerate \cite{bridle}. This
will modify the evolution of $\phi(z)$ and therefore the expected
evolution of the fine-structure constant between redshift $z=1$
and $z=0$. The details of the evolution of $\alpha_{EM}$ depend on
whether or not the moduli fields have a potential and if they
drive the apparent expansion at low redshift. Inferring the
evolution of $\alpha_{EM}$ as a function of redshift can therefore
give not only information about couplings between the moduli
field(s) but also about the potential \cite{Wetterich
1},\cite{Wetterich 2}. When a potential for the moduli field is
generated by an effective detuning of the brane tension, this
leads to a variation of the cosmological constant
\begin{equation}
\frac{\delta\Lambda}{\Lambda}=\frac{3}{8}\frac{\alpha}{\beta}\frac{\delta
\alpha_{EM}}{\alpha^2_{EM}},
\end{equation}
in the matter dominated era. For $\beta\approx \alpha$ this implies
that the variation of the cosmological constant is much smaller
than the cosmological constant itself.

Finally, we consider the evolution of the ratio $\mu = m_{p} / m_{e}$. If we
replace the variation of $\psi$ in terms of its cosmological
evolution we obtain from (\ref{eq: mu var})
\begin{eqnarray}
\frac{\delta \mu}{\mu} = \frac{1}{6} \, \frac{2 \pi}{ b^{SM}_{3}}
\frac{ 4 \alpha \beta}{1 + 2 \alpha^{2}}  \ln (1 + z).
\end{eqnarray}
Since $b^{SM}_{3}$ is negative, the variation of $\mu$ is
increasing in terms of $z$. This result does not agree with the
observations of Ivanchik {\it et al.} who have measured a
variation $\delta \mu / \mu = ( 5.02 \pm 1.82 ) \times 10 ^{-5}$
\cite{mp/me 1}-\cite{mp/me 2}. However, as claimed by this group,
a more conservative position should be adopted  until more
accurate measurements come. With this caveat, the observation of a
variation of $\mu$ can be taken to be
 $| \delta \mu / \mu | < 8 \times 10 ^{-5}$. In
terms of the brane moduli parameters, this constraint gives us
$|\alpha \beta| < 1 \times 10^{-4}$, which is compatible but
stronger than the above results. We would like to emphasize again
the fact that in the class of theories that we are presently
considering the sign of the variation $\delta \mu / \mu$ is the
same one of $\delta \alpha / \alpha$ (see equation (\ref{eq: mu
var})).

To summarize, in this paper we have presented two main results:
firstly, we have verified that the variation of coupling parameter
is frame-independent at the one-loop level in the quantum
correction.  Secondly, we have investigated the time variation of
coupling parameter in the fairly general brane-world model with
two boundary branes and bulk scalar field, whose low energy
dynamics is governed by two moduli fields \cite{US}. The existence
of these fields leads to corrections to general relativity. We
have argued that the only way to obtain time varying gauge
couplings is by coupling the gauge fields directly to the bulk
scalar field and allowing scalar field-dependent Yukawa couplings.
We have not considered the latter possibility and analyzed a
theory with two free parameters, $\alpha$ and $\beta$. The
parameter $\alpha$ defines the coupling of the brane tension to
the bulk scalar field while $\beta$ describes the dependence of
coupling parameter to the moduli field (see equation
(\ref{ansatz})). Interestingly, both parameters, although a priori
independent, have the same order of magnitude, when the current
constraints of variations of the fine structure constant and
current constraints by gravity experiments are considered. Our
work emphasizes once more the importance of experiments looking
for deviation to general relativity. Together with measurements of
the fine structure constant at high redshifts this would provide
useful constraints on theories beyond the standard model. In fact,
for $\alpha = \beta$ in the model discussed here, violations to
general relativity should be measurable very soon, if the
variation of $\alpha_{EM}$ reported in \cite{Webb 1}--\cite{Webb
3} is confirmed by future investigations.

\acknowledgements{ We are grateful to Fernando Quevedo,
Jean-Philippe Uzan and Thomas Dent for useful comments. We
acknowledge support from the Anglo-French alliance exchange
program. This work is supported in part by PPARC. G.A.P.
acknowledges the support of MIDEPLAN. P.B. is partially funded by
the RTN european programme HPRN-CT-2000-00148. }


\end{document}